\documentclass[a4paper, 11pt, oneside]{article}   	
\usepackage{geometry}                		
\usepackage{bm}
\usepackage{graphicx}				
\usepackage{authblk}								
								
\usepackage{amssymb}
\usepackage{amsmath}
\usepackage{lineno}
\usepackage[table]{xcolor}

\title{Eurasian cooling in response to Arctic sea--ice loss is not proved by maximum covariance analysis}
\author[1]{Giuseppe Zappa}
\author[1]{Theodore G.~Shepherd}
\author[2]{Paulo Ceppi}
\affil[1]{Department of Meteorology, University of Reading, Reading RG6 6BB, UK}
\affil[2]{Grantham Institute for Climate Change and the Environment, Imperial College, London SW7 2AZ, UK}
\date{}							
\begin{document}
\maketitle

The extent to which the ongoing decline in Arctic sea ice affects mid--latitude climate has received great attention and polarised opinions. The basic issue is whether the inter-annual variability in Arctic sea ice is the cause of, or the response to, variability in mid-latitude atmospheric circulation \cite{Shepherd2016}. A recent paper by Mori et al.~(M19, \cite{Mori2019}) claims to have reconciled previous conflicting studies by showing that a consistent mid--latitude climate response to inter--annual sea--ice anomalies can be identified in both the ERA-Interim reanalysis, taken as observations, and in an ensemble of atmosphere--only (AMIP) climate model simulations. We here demonstrate that such a conclusion cannot be drawn, due to issues with the interpretation of the maximum covariance analysis performed. After applying the M19 approach to the output from a simple statistical model, we conclude that a predominant atmospheric forcing of the sea--ice variability, rather than the converse, is the most plausible explanation of the results presented in M19. 

A leading mode of internal atmospheric variability is associated, in its positive phase, with a Siberian anticyclone, a Warm Arctic and Cold Eurasia (WACE mode, Fig S4 in M19). It is debated whether anomalies in the extent of Barents and Kara sea ice can modulate the frequency of occurrence of this mode, given that the Siberian circulation anomaly  could  itself force sea--ice anomalies by warming the Arctic region. To discriminate between these two possible scenarios, M19 rely on identifying a mode of year--to--year co--variability in the winter--mean (DJF) Eurasian surface temperature between the ERA--Interim reanalysis and an ensemble of AMIP simulations, i.e.~climate runs forced by observed oceanic conditions. The approach is well designed: if sea ice forces circulation, the WACE modes in the ERA--Interim and AMIP simulations, including their Arctic--Eurasian temperature dipoles, should covary in time. If instead sea ice responds to circulation, the WACE modes should not covary:  in the real world the WACE mode would force sea--ice variability, but sea ice would at most force a monopole of Arctic temperature variability, via local thermodynamic processes, in the AMIP simulations. Covarying signals can also result from the forcing due to sea surface temperature (SST) variability, whose role can be assessed with dedicated experiments.

The leading mode of Eurasian surface temperature co--variability between ERA--Interim and the AMIP simulations is identified in M19 via maximum covariance analysis (MCA), as implemented through the singular value decomposition (SVD) of the co--variance matrix between the two surface temperature fields in 0E--180E, 20N--90N. However, the pair of singular vectors that comprise the co--varying mode are not displayed in the paper. Instead, the authors discuss the mode in terms of the homogeneous regression maps obtained by regressing each field on the expansion coefficient (EC) of its own singular vector (see SI for an overview of the methodology). The homogeneous regression maps are not necessarily directly related to the singular vectors, and hence to the structure of the co--varying mode \cite{Bretherton1992,Wallace1992}. This is because the ECs are obtained by projecting the two analysed fields on their own singular vectors. Hence, in addition to reflecting the co--varying mode, the ECs include variance generated by any internal mode of variability that is not orthogonal to the singular vectors themselves. When the original fields are regressed on their own ECs, such internal modes can be aliased into the homogeneous regression maps. The way to isolate the structure of the co--varying mode is via heterogeneous regression maps, in which each field is regressed on the EC from the other field \cite{Bretherton1992}.

The potential pitfalls of solely examining homogeneous maps are explored by applying the statistical method from M19 to the output from a simple statistical model that qualitatively incorporates the influence of sea ice and of the WACE mode on surface temperature variability. In the simple model, the direction of the interaction between atmospheric circulation and sea ice can be directly controlled (see SI). Regardless of whether sea--ice variability forces atmospheric circulation or vice--versa, we find that the homogeneous regressions for the leading co--varying mode always show WACE--like patterns characterised by a warm Arctic, a cold Eurasia and a positive Siberian surface pressure anomaly. The same is not true for the heterogeneous regressions, which correctly identify distinct pairs of co--varying patterns - either WACE--like or Arctic--monopole like - depending on the presence and direction of the interaction between sea--ice and the atmospheric circulation (Table \ref{tab:results}). Because they alias in the internal variability in the WACE mode, homogeneous regressions are insufficient to discriminate between these different scenarios from the simple model. It is also dangerous to interpret the correlation between the sea ice and the expansion coefficients as the fraction of WACE variability forced by sea--ice, as in M19. An anti--correlation is always to be expected provided that sea ice can affect Arctic temperatures via local thermodynamic processes (Table \ref{tab:results}).

\begin{figure*}
\centering
 \includegraphics[width=0.7\textwidth]{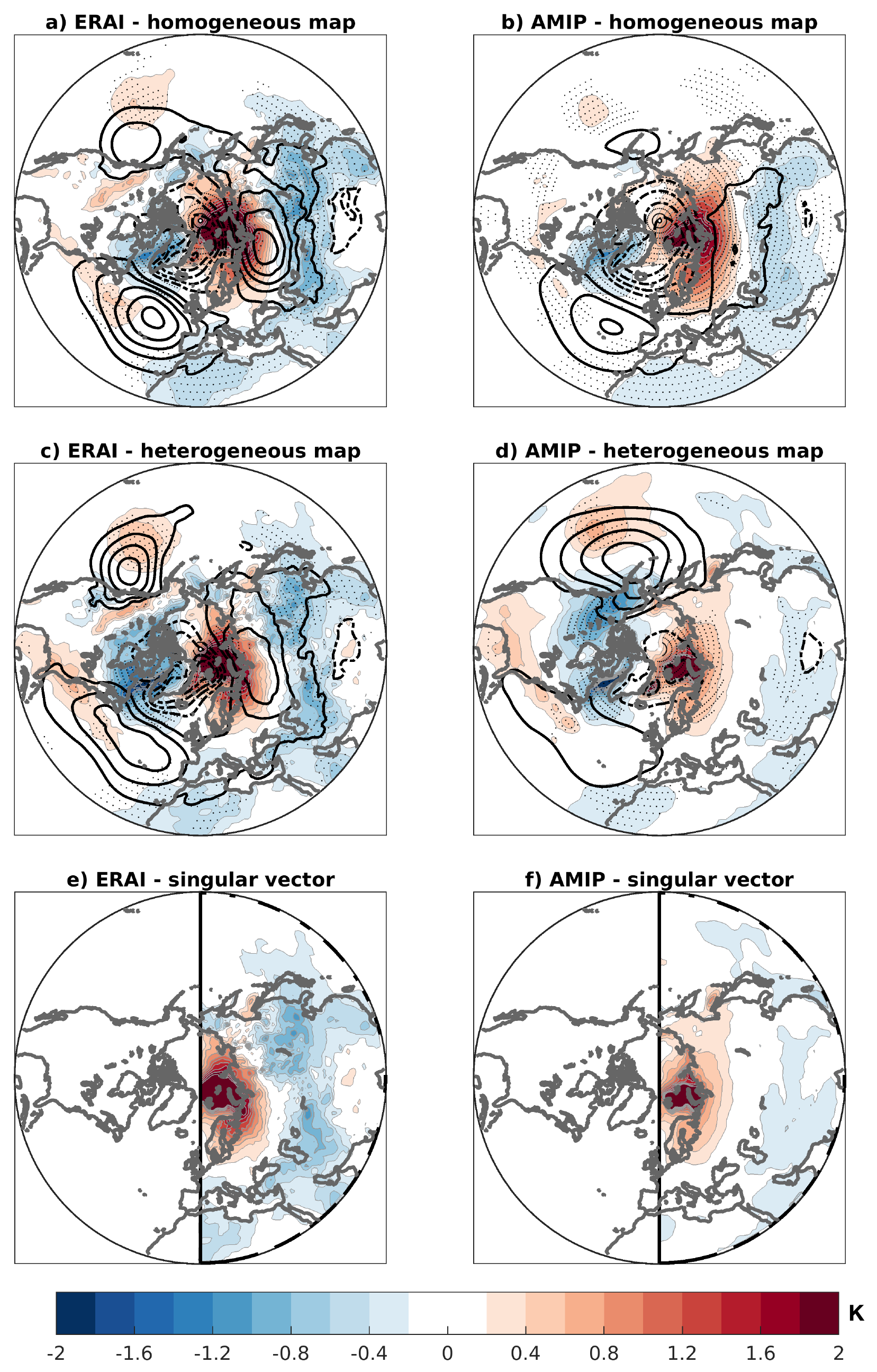}
\caption{Comparison between a--b) the homogeneous regression maps, c--d) the heterogeneous regression maps and e--f) the singular vectors following the analysis in M19. All maps are scaled to one standard deviation anomaly in the expansion coefficients (see the SI for details). a, c and e refer to the ERA--Interim reanalysis, while b, d and f refer to the AMIP simulations. Shading shows the near--surface atmospheric temperature (K), and the contours the sea--level pressure with a c.i.~of 0.5 hPa, solid for positive and dashed for negative. Stippling indicates statistical significance in the regression coefficients at the 5\% level, as obtained by bootstrapping the individual years with resampling.}
\label{fig:hetero}       
\end{figure*}

We therefore compare the structure of the co--varying mode between the ERA--Interim and AMIP simulations obtained from the homogeneous  (Fig.~\ref{fig:hetero}a-b) and heterogeneous (Fig.~\ref{fig:hetero}c-d) regression maps. The only difference with M19 is that, since the MIROC4 simulations are unavailable to the authors, the AMIP multi--model ensemble consists of 6 rather than 7 models. Nonetheless, the homogeneous maps  bear a strong resemblance to those presented in M19, featuring the WACE--mode--like temperature dipole between the Arctic and Central Eurasia together with Siberian sea level pressure anomalies. These features are also present, albeit with a weaker amplitude, in the heterogeneous map from ERA--Interim, but not in the heterogeneous map from the AMIP ensemble. In particular, there is no evidence of a sea level pressure signal in Eurasia, as well as no evidence of a cold anomaly in Central Eurasia, though very weak temperature anomalies  persist further south. Only the heterogeneous maps correctly reproduce the co--varying temperature signals identified by the singular vectors (Fig.~\ref{fig:hetero}e-f) within their domain of definition in the Eastern Hemisphere. 

The covariance of a WACE--mode like dipole in ERA-Interim with an Arctic temperature monopole in the AMIP simulations is most consistent with an atmospheric forcing of sea ice variability, as suggested by the statistical model. This could not be concluded by solely inspecting the homogeneous regression maps presented in M19 due to the aliasing of the internal variability associated with the WACE mode. The heterogenous maps still reveal covarying pressure signals in the North Pacific and North Atlantic but their nature, and particularly the role played by SSTs and sea ice in forcing them, remains to be investigated. In conclusion, we suggest that whilst the MCA proposed in M19 is useful, a more plausible interpretation of their results is that it is the WACE mode that predominantly forces sea--ice variability, in contrast with the claim of M19. Of course, this relies on the realism of the models.

%

\definecolor{lightlightgray}{gray}{0.95}
\newcommand{\g}{\cellcolor{lightlightgray}}

\definecolor{lightgray}{gray}{0.75}
\newcommand{\G}{\cellcolor{lightgray}}

\begin{table}[htp]
\begin{center}
\begin{tabular}{|c||cc||cc||cc||}
\hline
& \multicolumn{6}{c||}{Homogeneous regression}\\
\hline
& \multicolumn{2}{c||} {Ice $\rightarrow$ Atm} & 
\multicolumn{2}{c||}{Ice $\nsim$ Atm} &
\multicolumn{2}{c||}{Ice $\leftarrow$ Atm}\\
\hline
& World & AMIP & World & AMIP & World & AMIP \\
\hline
 {$T_{Arctic}$} & \g 2.2  & \g 2.2  & \g 1.4   & \g 1.4   & \g 2.2 & \g 1.7\\
 {$C$} 		      & \g 1.4     & \g 1.4 & \g 0.7   & \g 0.7   & \g 0.9   & \g 0.6 \\
 {$T_{Asia}$}   & \g -1.4 & \g -1.4 & \g  -0.7 & \g  -0.7 & \g -0.9 & \g -0.6\\
 \hline
 & \multicolumn{6}{c||}{Heterogeneous regression}\\
 \hline
 {$T_{Arctic}$} & \g 2.3 & \g 2.3  & \G 1.4  &  \G1.4 & \g  2.3 &  \G 1.7\\
 {$C$} 		      & \g 1.2 & \g 1.2  & \G 0.0   &  \G0.0 & \g 0.8  & \G 0.0\\
 {$T_{Asia}$}   & \g-1.2 & \g-1.2 &  \G 0.0  & \G 0.0 & \g -0.8 & \G 0.0\\
 \hline
 & \multicolumn{6}{c||}{Correlation between sea ice (I) and ECs}\\
 \hline
 {$r_{I,EC}$} & -0.86 & -0.86 & -0.71 & -0.71 & -0.93 & -0.82\\
\hline
\end{tabular}
\end{center}
\caption{The application of the M19 approach to the output from a simple statistical model (see SI for details) using homogeneous (top) and heterogeneous (bottom) regressions. The modes of co--variability between the ``real--world" (World) and ``AMIP--world" (AMIP) systems are described in terms of the following variables from the simple model: $T_{Arctic}$ (representing Arctic temperature), $C$ (atmospheric circulation, positive for a Siberian anticyclone) and $T_{Asia}$ (Central Eurasian temperature). Light grey cells indicate WACE--like patterns, and dark grey cells indicate Arctic temperature monopole patterns. The columns report results from different setups of the statistical model: sea ice driving the circulation (Ice $\rightarrow$ Atm), no interaction between sea ice and circulation (Ice $\nsim$ Atm), and circulation driving the sea ice (Ice $\leftarrow$ Atm). The bottom line reports the correlation between the variability in the sea ice (I) and that in the expansion coefficients.}
\label{tab:results}
\end{table}%

\bibliographystyle{ieeetr}
\bibliography{biblio}

\begin{thebibliography}{1}

\bibitem{Shepherd2016}
T.~G. Shepherd, ``{Effects of a warming Arctic},'' {\em Science}, vol.~353,
  pp.~989--990, 2016.

\bibitem{Mori2019}
M.~Mori, Y.~Kosaka, M.~Watanabe, H.~Nakamura, and M.~Kimoto, ``{A reconciled
  estimate of the influence of Arctic sea-ice loss on recent Eurasian
  cooling},'' {\em Nat Clim Change}, vol.~9, pp.~123--129, 2019.

\bibitem{Bretherton1992}
C.~S. Bretherton, C.~Smith, and J.~M. Wallace, ``{An Intercomparison of Methods
  for Finding Coupled Patterns in Climate data},'' {\em J Climate}, vol.~5,
  pp.~541--560, 1992.

\bibitem{Wallace1992}
J.~M. Wallace, C.~Smith, and C.~S. Bretherton, ``{Singular value decomposition
  of wintertime sea surface temperature and 500-mb height anomalies},'' {\em J
  Climate}, vol.~5, pp.~561--576, 1992.

\end{thebibliography}


\begin{thebibliography}{1}

\bibitem{Bretherton1992}
C.~S. Bretherton, C.~Smith, and J.~M. Wallace, ``{An Intercomparison of Methods
  for Finding Coupled Patterns in Climate data},'' {\em J Climate}, vol.~5,
  pp.~541--560, 1992.

\end{thebibliography}

\end{document}


\maketitle


\section{MCA: a brief overview}
Maximum covariance analysis (MCA) enables to identify pairs of spatial patterns that covary in time between two fields ($A$ and $B$) and that explain as much as possible of the covariance between the two. $A$ and $B$ can have different spatial dimensions ($x$), but are restricted to having the same temporal dimension ($t$). Without loss of generality, the fields $A$ and $B$ can be described as two bi--dimensional matrices having dimensions, respectively, $X_a \times T$ and $X_b \times T$. A set of pairs of co--varying spatial patterns is then identified by decomposing the cross--covariance matrix (C) between $A$ and $B$ via its singular value decomposition:
\begin{equation}
C=\frac{1}{T}AB^{T}=U\Sigma V^{T}=\sum_k{\sigma_k \cdot \bm{u_k}\,\bm{v_k}^T}\\
\end{equation}
where $U=(\bm{u_1},\dots,\bm{u_N})$ and $V=(\bm{v}_1,\dots,\bm{v_N})$ are squared matrices containing the so-called left ($\bm{u_k}$) and right singular vectors ($\bm{v_k}$) in their columns. For any value of $k$, each pair of left and right singular vectors describe a mode of co--variability between $A$ and $B$. $\Sigma$ is a diagonal rectangular matrix, with the values on the diagonal ($\sigma_k$) reflecting the magnitude of the squared covariance between $A$ and $B$ explained by each mode. Each successive mode following $k=1$ explains a smaller fraction of squared covariance.

The time variability associated with a co--varying mode can be described in terms of \emph{expansion coefficients (ECs)} ($a_k$ and $b_k$) obtained by projecting the original fields on their own singular vectors, i.e.~$a_k=\bm{u_k}^T\,A$ and $b_k=\bm{v_k}^T\,B$. Since the ECs are computed via a projection, the expansion coefficient $a_k$ also incorporates variance from any internal mode of variability in $A$ that is not orthogonal to the singular vector $u_k$. In the same way, the expansion coefficient $b_k$  includes variance due to any mode of internal variability in $B$ not orthogonal to $v_k$. The presence of such internal variance in $a_k$ and $b_k$ has implications for the interpretation of the spatial patterns associated with the expansion coefficients via linear regression. In particular, the patterns obtained by regressing each original field onto their own expansion coefficients (called homogeneous maps) tend to reflect both the structure of the co--varying mode and of any internal modes that contribute to the variance of $a_k$ and $b_k$. As a result, homogeneous regression maps do not necessarily have a direct relationship to the singular vectors, which exclusively contain the structure of the co--varying mode. After appropriate scaling (see below), the structure of the co--varying mode can instead be recovered via heterogeneous regression maps, so--called because they are obtained by regressing each field on the expansion coefficient of the other field \cite{Bretherton1992}.

\section{Scaling of singular vectors and regression maps}
In Fig 1 of the commentary, the homogeneous and heterogeneous regression maps, as well as the singular vectors, are scaled to correspond to a one standard deviation anomaly in the ECs. In particular, the scaled ($\,\tilde{}\,$) expansion coefficients and singular vectors are computed as:
\begin{equation}
\begin{aligned}
\tilde{a}_k&=a_k/\sigma_{a_k}\\
\tilde{b}_k&=b_k/\sigma_{b_k}\\
\tilde{\bm{u}}_k&=\bm{u_k}\cdot \sigma_{a_k}\\
\tilde{\bm{v}}_k&=\bm{v_k}\cdot \sigma_{b_k},\\
\end{aligned}
\end{equation}
where $\sigma_{a_k}$ and $\sigma_{b_k}$  are the standard deviations of the respective ECs. For each grid point, the homogeneous regression maps are obtained as the slope ($m$) of the linear regressions passing through zero, between each field and their own scaled expansion coefficients. Using the notation $m=<y,x>$ to indicate the slope of y regressed on x and the standard formulas for least square regression: 
\begin{equation}
\begin{aligned}
<A,\tilde{a}_k>&=\frac{A\tilde{a}_k^T}{\tilde{a}_k \tilde{a}_k^T},\quad\textrm{homogeneous map for A}\\
<B,\tilde{b}_k>&=\frac{B\tilde{b}_k^T}{\tilde{b}_k \tilde{b}_k^T},\quad\textrm{homogeneous map for B}
\end{aligned}
\end{equation}

The heterogeneous regression maps can be interpreted in a similar way, but the linear regressions are computed conditional on the  EC from the other field:
\begin{equation}
\begin{aligned}
<A,E(\tilde{a}_k | \tilde{b}_k)>&=\frac{A\tilde{b}_k^T}{\tilde{b}_k \tilde{b}_k^T} \cdot \frac{1}{r(a_k,b_k)},\quad\textrm{heterogeneous map for A}\\
<B, E(\tilde{b}_k | \tilde{a}_k)>&=\frac{B\tilde{a}_k^T}{\tilde{a}_k \tilde{a}_k^T} \cdot \frac{1}{r(a_k,b_k)},\quad\textrm{heterogeneous map for B}\\
\end{aligned}
\end{equation}
$r(a_k,b_k)$ is the correlation between the two expansion coefficients, and it represents the scaling factor needed to make the scaled heterogeneous regressions equivalent to the scaled singular vectors. 

\section{A simple model of Arctic--midlatitude interactions}
The ability of the approach from M19 to identify climate impacts from sea--ice variability is tested on the output from a simple statistical model of Arctic and Eurasian temperature variability, in which the role played by sea ice can be directly controlled. Despite its simplicity, the statistical model, which is defined as follows, incorporates the key role of sea ice and circulation to force the temperature anomalies discussed in M19:
\begin{equation*}
\begin{aligned}
&T_{Arctic}&&=-\alpha \, I&&+\beta \,C +   \epsilon_{Arctic} \\
&T_{Asia}&&=0&&-\beta \, C+ \epsilon_{Asia}  \\
&\hat{T}_{Arctic}&&=-\hat{\alpha} \, I&&+\hat{\beta} \,\hat{C} +   \hat{\epsilon}_{Arctic} \\
&\hat{T}_{Asia}&&=0&&-\hat{\beta} \, \hat{C}+ \hat{\epsilon}_{Asia}
\end{aligned}
\begin{aligned}
&\left.\vphantom{\begin{aligned}
&T_{Arctic}&&=-\alpha \, I&&+\beta \,C +   \epsilon_{Arctic} \\
&T_{Asia}&&=0&&-\beta \, C+ \epsilon_{Asia}  \\
  \end{aligned}}\right\rbrace\quad\text{Real world}\\
&\left.\vphantom{\begin{aligned}
&\hat{T}_{Arctic}&&=-\hat{\alpha} \, I&&+\hat{\beta} \,\hat{C} +   \hat{\epsilon}_{Arctic} \\
&\hat{T}_{Asia}&&=0&&-\hat{\beta} \, \hat{C}+ \hat{\epsilon}_{Asia},  
  \end{aligned}}\right\rbrace\quad\text{AMIP world}
\end{aligned}
\end{equation*}
$T_{Arctic}$ and $T_{Asia}$ represent, respectively, the surface temperature anomalies in the Arctic and Central Eurasia, $I$ represents the sea--ice anomaly in the Barents and Kara region and $C$ represents the atmospheric circulation anomaly associated with the WACE mode, so that a positive value in $C$ corresponds to an anticyclonic circulation anomaly in Siberia.  $\alpha$ and $\beta$ are parameters quantifying the respective influence of sea ice and circulation anomalies on temperature anomalies. For simplicity, it is here assumed that circulation anomalies drive  temperature anomalies of equal amplitude, but opposite sign, between the Arctic and Eurasia. $\epsilon$ represents additional variability in the surface temperature, here modelled as Gaussian white noise, that is induced by other local processes or by other atmospheric processes than the WACE mode. The variables representing the output from AMIP simulations are denoted with a $\hat{}$. The distinction between the real--world and AMIP--world equations comes from the asymmetry in the system introduced by $I$, as it is the real--world sea ice that affects Arctic surface temperature in both the real world and the AMIP world. 

Different possible setups are considered based on the interaction between sea ice  and circulation anomalies. First, to reflect the conclusions from M19, we consider a setup in which sea ice drives the probability of occurrence of the WACE mode. In this setup, we define $I$ as an independent random variable, and circulation (both $C$ and $\hat{C}$) as a random variable whose mean expected value depends on sea ice: 
\begin{equation*}
\left.
\begin{aligned}
&I&&\overset{iid}{\sim}&&N(0,1)&&\\
&C|I&&\sim&&N(-\gamma\cdot I,1)&&\\
&\hat{C}|I&&\sim&&N(-\hat{\gamma}\cdot I,1),&&
\end{aligned} \right\rbrace\quad\text{Sea ice drives circulation}
\end{equation*}
where $N(0,1)$ is a normal distribution of zero mean and unit variance, and $\gamma$ and $\hat{\gamma}$ are positive parameters describing the strength of the interaction in the real world and in the AMIP world, respectively. The minus sign before $\gamma$ reflects the conclusion from M19 that a negative anomaly in the sea ice can force a positive WACE mode. This simple model does not include the possibility that circulation or surface temperature are forced by the variability in the SSTs. This is acceptable for the purpose of this note since M19 suggest that SSTs play only a minor role in the forcing of the WACE mode. 

As a contrasting case, we consider an opposite setup in which it is circulation that drives sea--ice variability, rather than vice--versa. In this case, $C$ -- as well as $\hat{C}$ -- is modelled as an independent random process, and $I$ as a random variable whose mean value depends on $C$ in the real world:
\begin{equation*}
\left.
\begin{aligned}
&C&&\overset{iid}{\sim}&&N(0,1)&&\\
&\hat{C}&&\overset{iid}{\sim}&&N(0,1)&&\\
&I|C&&\sim&&N(-\gamma\cdot C,1).&&
\end{aligned} \right\rbrace\quad\text{Circulation drives sea ice}
\end{equation*}
As in the previous setup,  $\gamma$ is defined to be positive and it describes the strength of the interaction. The minus sign before $\gamma$ implies that a positive anomaly in C, i.e. a Siberian anticyclone, drives a reduction in sea ice, for example by advecting warm air into the Arctic. The two setups described above  become identical for $\gamma=0$, which describes a third setup in which there is no interaction between the variability in the circulation and in the sea ice. 

For any given parameter setting, the statistical model is used to generate two sets of synthetic time series of $10^6$ values in the variables $T_{Arctic}$, $\hat{T}_{Arctic}$, $T_{Asia}$, $\hat{T}_{Asia}$, $C$, $\hat{C}$ and $I$. The same MCA adopted in M19 is applied to the two 2--dimensional time series $(T_{Arctic},T_{Asia})$ and $(\hat{T}_{Arctic},\hat{T}_{Asia})$ in order to test whether the approach is able to discriminate between cases with different values of $\gamma$ and with different directions of interaction between circulation and sea ice.  In particular, three setups of the model's parameters are considered and discussed in Table 1 of the commentary:  one in which sea ice drives circulation (with $\gamma=1$),  one in which there is no interaction between the sea ice and the circulation anomalies ($\gamma=0$), and one in which it is the atmosphere that drives the sea ice (also with $\gamma=1$). For simplicity, a no--bias configuration has been chosen for the other parameters in the statistical model, i.e. $\hat{\gamma}=\gamma$, $\hat{\alpha}=\alpha=1$ and $\hat{\beta}=\beta=1$. 

\bibliographystyle{ieeetr}
\bibliography{biblio}